\documentclass[twocolumn,english,rsi,superscriptaddress]{revtex4-1}
\usepackage[T1]{fontenc}
\usepackage[latin9]{inputenc}
\usepackage{graphicx}

\makeatletter
 \@ifundefined{textcolor}{}
 {%
  \definecolor{BLACK}{gray}{0}
  \definecolor{WHITE}{gray}{1}
  \definecolor{RED}{rgb}{1,0,0}
  \definecolor{GREEN}{rgb}{0,1,0}
  \definecolor{BLUE}{rgb}{0,0,1}
  \definecolor{CYAN}{cmyk}{1,0,0,0}
  \definecolor{MAGENTA}{cmyk}{0,1,0,0}
  \definecolor{YELLOW}{cmyk}{0,0,1,0}
  }

\usepackage{babel}

\makeatother

\usepackage{babel}
\begin{document}

\title{Multichannel photodiode detector for ultrafast optical spectroscopy}

\author{T. Mertelj }

\email{tomaz.mertelj@ijs.si}

\selectlanguage{english}%

\affiliation{Complex Matter Dept., Jozef Stefan Institute, Jamova 39, Ljubljana,
SI-1000, Ljubljana, Slovenia }

\author{N.Vuji\v{c}i\'{c} }

\affiliation{Complex Matter Dept., Jozef Stefan Institute, Jamova 39, Ljubljana,
SI-1000, Ljubljana, Slovenia }

\affiliation{Institute of Physics, Bijeni\v{c}ka 46, HR-10000 Zagreb, Croatia}

\author{T. Borzda }

\affiliation{Complex Matter Dept., Jozef Stefan Institute, Jamova 39, Ljubljana,
SI-1000, Ljubljana, Slovenia }

\affiliation{Jožef Stefan International Postgraduate School, Jamova 39, SI-1000
Ljubljana, Slovenia}

\author{I. Vaskivskyi }

\affiliation{Complex Matter Dept., Jozef Stefan Institute, Jamova 39, Ljubljana,
SI-1000, Ljubljana, Slovenia }

\author{A. Pogrebna }

\affiliation{Complex Matter Dept., Jozef Stefan Institute, Jamova 39, Ljubljana,
SI-1000, Ljubljana, Slovenia }

\affiliation{Jožef Stefan International Postgraduate School, Jamova 39, SI-1000
Ljubljana, Slovenia}

\author{D. Mihailovic}

\affiliation{Complex Matter Dept., Jozef Stefan Institute, Jamova 39, Ljubljana,
SI-1000, Ljubljana, Slovenia }

\affiliation{Jožef Stefan International Postgraduate School, Jamova 39, SI-1000
Ljubljana, Slovenia}

\affiliation{CENN Nanocenter, Jamova 39, Ljubljana SI-1000, Slovenia}

\date{\today}
\begin{abstract}
Construction and characterization of a multichannel photodiode detector
based on commercially available components with high signal to noise
of $\sim10^{6}$ and a rapid frame rate, suitable for time resolved
femtosecond spectroscopy with high repetition femtosecond sources,
is presented. 
\end{abstract}
\maketitle

\section{introduction}

In recent years femtosecond spectroscopy has been established as a
standard tool to investigate physics of novel solid-state materials
such as nonconventional superconductors, electronic crystals and related
collectively ordered systems.\cite{ChekalinFartzdinov1991,DemsarPodobnik1999,DemsarBiljakovic99,LobadTaylor2000,KaindlWoerner2000,DvorsekKabanov2002,DemsarForro2002,SegreGedik2002,TomimotoMiyasaka2003,KaindlCarnahan2005,GedikOrenstein2003,ParsenkumarOkamura2005,YusupovMertelj2008,LiuToda2008,GiannettiCoslovich2009,MerteljOslak2009,MerteljKabanov2009prl,SchaeferKabanov2010,TorchinskyChen2010,GadermaierAlexandrov2010,MansartBoschetto2010,KusarMertelj2011,GiannettiCilento2011,KimPashkin2012,MansartCottet2012}
Majority of the published work was done using narrowband optical probe
pulses. There are several reasons for use of narrowband probes. Apart
from simplicity of a setup, the main reason is the necessity to use
lockin modulation techniques to reject the laser noise in order to
detect small changes in optical responses. On the other hand, use
of broadband optical pulses\cite{GadermaierAlexandrov2010,GiannettiCilento2011,PogrebnaVujicic2014,KimPashkin2012,MansartCottet2012}
can provide additional spectral information which is time consuming
to obtain by means of wavelength-tunable narrowband optical pulses
\cite{OgasawaraKimura2001,TomimotoMiyasaka2003,OkamotoMiyagoe2010}. 

Since many low temperature ordered states are easily destroyed by
rather weak photoexcitation.\cite{OgasawaraKimura2001,KusarKabanov2008,GiannettiCoslovich2009,YusupovMertelj2010}
the energy of the probe pulses must be kept low enough. In the case
of broadband probe pulses this requires very efficient multichannel
detection system that is fast enough to enable employment of the lockin-like
modulation techniques\cite{PolliLuer2007,SmithLight2010} to suppress
the laser noise. Such systems have been implemented recently in a
kHz system\cite{PolliLuer2007} and cavity dumped fast repetition
system\cite{CilentoGianneti2010} with the signal to noise ratio (SNR)
$\sim10^{4}$ enabling detection of $>10^{-4}$ photoinduced relative
changes of optical constants.

Here we present construction and characterization of a multichannel
photodiode detection system built from commercially available components
with $\mathrm{SNR}>10^{5}-10^{6}$ enabling lockin-like modulation
techniques with modulation frequency up to 3 kHz appropriate for experiments
using high repetition ($\gg1$ kHz) pulsed laser sources.

\section{Statement of the problem}

In time resolved pump-probe experiments a relatively small changes
of the reflectivity or transmittance of the sample are measured, 
\begin{equation}
\Delta G_{j}/G_{j}=\Delta I_{j}/I_{j},\label{eq:dRoveR}
\end{equation}
where $G$ stands either for reflectivity, $R_{j}$, or transmittance,
$T_{j}$, and $I_{j}$ represents the light intensity in the channel
$j$. In a photodetector the SNR of a single channel is fundamentally
limited by the number of acquired photoelectrons, $n_{\mathrm{pe}}$.
For $\Delta G_{j}/G_{j}\sim10^{-6}$ the $\mathrm{SNR}=10^{6}$ is
required. In the shot-noise limit $n\mathrm{_{pe}}=10^{12}$ photoelectrons
need to be acquired corresponding to a charge of $q=160$ nC. Assuming
100\% quantum efficiency, this corresponds to $\sim250$ nJ of optical
energy (at 800-nm photon wavelength) per channel. Taking a typical
convenient laser spot size%
\footnote{The spot size is often limited by the sample size.%
} of $\sim(100$$\mu$m$){}^{2}$ this leads to $\sim2.$5 mJ/cm$^{2}$
of reflected/transmitted optical energy per channel. Limiting the
single probe pulse fluence to 10 $\mu$J/cm$^{2}$ and assuming 50
channels the required number of probe pulses is $\sim10000$ resulting
in the minimum $\sim10$ s acquisition time per spectrum for a kHz
laser system. With high repetition lasers this time can be considerably
shorter, below one second. The $\mathrm{SNR}=10^{6}$ is therefore
feasible, provided that noise from other sources, especially the laser
source noise, can be eliminated.

\begin{figure}
\includegraphics[angle=-90,width=1\columnwidth]{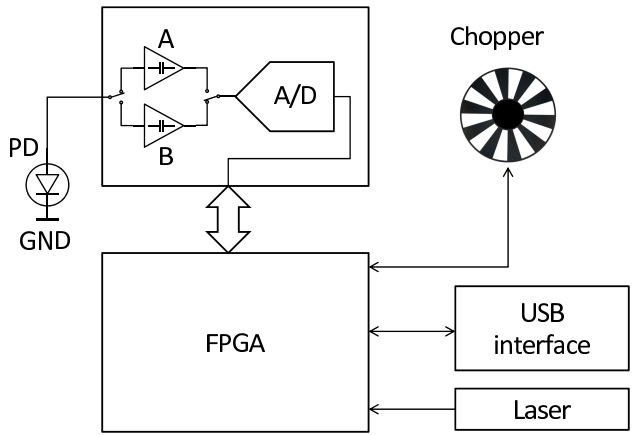}

\caption{A schematic diagram of the detector. PD - photodiode (one channel),
GND - ground, A, B - integrators. A/D - analog to digital converter,
FPGA - fully programmable gate array.}
\label{fig:schematic}
\end{figure}

Modern spectroscopic CCDs have output register well capacity of $\sim10^{6}$
($q\sim0.16$ pC) electrons and can achieve $\sim1$ kHz readout rate
(see for example ANDOR datasheets) leading to impractical%
\footnote{To measure time evolution spectra at different delays usually requires
several tens of spectra per scan. %
} accumulation times of%
\footnote{Here we assume that by a proper optical setup $\sim10$ columns can
be joined into a single spectral channel leading to an effective $\sim100$
channel detector.%
} $>100$ s per spectrum. Commercially available CMOS and NMOS linear
image sensors, on the other hand, offer on-chip effective well capacities
in excess of 10 pC (see for example Hammamatsu data sheets), but mainly
offer sequential readout only, which introduces dead-time and exposure-control
problems%
\footnote{With a kHz laser a single optical pulse per scan, which is well defined
in time, implements its own shutter. With high repetition laser an
external shutter is needed to prevent exposure during the readout.%
} when used in conjunction with high repetition lasers.

Recently, a custom CMOS detector was presented,\cite{SmithLight2010}
incorporating four independent storage capacitors per pixel with the
effective well capacity of $600\cdot10^{6}$ ($q\sim100$ pC) photoelectrons,
capable of operating with up to $\sim40$ kHz frame rate, enabling
a multichannel pseudo-lockin detection at up to $\sim10$ kHz modulation
frequency. To our best knowledge this detector is not yet commercially
available so a similar solution based on commercially available components
is desirable.

\section{Our solution}

Our solution is based on a multichannel Texas Instruments DDC-series
current-input analog-to-digital (A/D) converter, that offers up to
150 pC effective well capacity with up to 6.25 kHz frame rate. Thus,
$n_{\mathrm{pe}}=10{}^{12}$ photoelectrons can be accumulated on
a timescale of second. Due to the dual switched integrator front-end
the integration window of each input is synchronous, enabling precise
exposure control with almost no dead time, the condition which is
required in the case of high repetition lasers. 

Different types of photodiode arrays can be coupled to the AD converter.
We chose a 46-channel Hamamatsu S4111 series silicon photodiode array
and a 16-channel Hamamatsu G7150-16 InGaAs PIN photodiode array for
visible and near infrared spectral regions, respectively. The rather
small number of channels is not a disadvantage since the high SNR
requirement together with the probe pulse energy limitation inherently
constrain the channel number. 

The digital interface to the AD converter (see Fig. \ref{fig:schematic})
was implemented in a fi{}eld programmable gate array (FPGA) providing
control logic, clocks, synchronization with a chopper and laser, frame
averaging as well as data transfer to an USB interface circuit. 

Since for transient reflectivity/transmittance measurements a full
phase sensitive detection is not necessary two spectra are acquired
during each pump modulation period. The photoinduced intensity is
calculated by,

\begin{equation}
\Delta I_{j}/I_{j}=\frac{I_{j,\mathrm{on}}-I_{j,\mathrm{off}}}{I_{j,\mathrm{off}}},
\end{equation}
where $I_{j,\mathrm{on}}$ and $I_{j,\mathrm{off}}$ correspond to
intensities measured during opening and blocking periods of the chopped
pump beam, respectively. To avoid exposure during the integrator switching
time the detector and chopper are synchronized to the laser source
by the FPGA circuitry. %
\footnote{For the cases presented here a high repetition (250 kHz) optical regenerative
amplifier (Coherent REGA) was employed.%
}

\section{Results}

\begin{figure}
\textbf{\vspace{4mm}
\includegraphics[angle=-90,width=1\columnwidth]{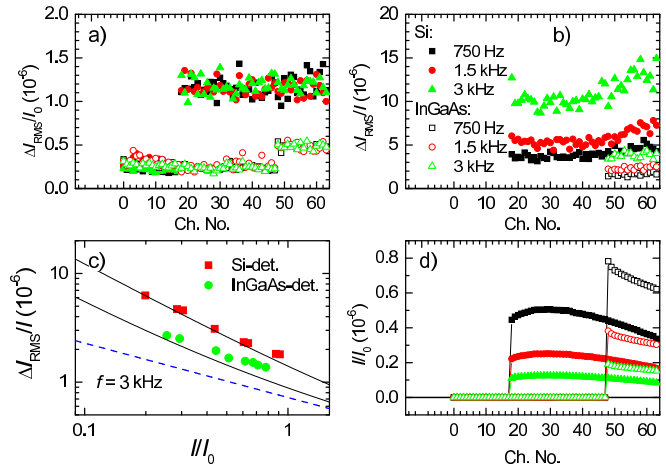}}\caption{(a) The non-illuminated-detector noise floor at different modulation
frequencies. (b) The signal-noise floor measured under CW halogen-lamp
illumination. The corresponding intensities are shown in (d). The
non uniformity is due to nonhomogenoeus illumination. (c) The CW-illumination
noise floor as a function of intensity. The blue dashed line is the
shot-noise floor while the thin black lines are the calculated total
noise taking into account both, the dark noise from (a) and the shot
noise, with no fitting parameters. At each measurement 2000 samples
were averaged, corresponding to the total available photoelectron
charge capacity per channel of, $q_{\mathrm{max}}=300$ nC. $I_{0}$
is the saturation intensity at the full effective-well capacity of
$q_{0}=150$ pC per sample. Only 48 channels (ch. no. 19 to 63) in
the case of Si photodiode array and 16 channels (ch. no. 49 to 63)
in the case of InGaAs photodiode array were connected.}
\label{fig:noise}
\end{figure}

In Fig. \ref{fig:noise} we show results of a basic characterization
of the detector. In addition to the photoelectron shot noise the integrators
input noise, the photodiode-capacitance thermal noise and other electronic
noise picked from environment are present, while the digitization
noise should be negligible with respect to the shot-noise due to the
20-bit resolution of the on-chip ADCs. 

In Fig. \ref{fig:noise} (a) we show the dark noise of the detector
measured under typical experimental conditions by averaging 2000 readouts
corresponding to $0.6-2$ s total integration time per single spectrum.
The unconnected channels%
\footnote{Channels 0 to 18 for the Si detector and 0 to 48 for InGaAs detector.%
} show noise of $\sim3\cdot10^{-7}I_{0}$, where $I_{0}$ corresponds
to the saturation intensity at the full effective-well capacity of
$q_{0}=150$ pC. The connected channels show additional noise of $\sim1.2\cdot10^{-6}I_{0}$
and $\sim0.5\cdot10^{-6}I_{0}$ for the Si and InGaAs detectors, respectively.

\begin{figure}
\includegraphics[angle=-90,origin=lB,width=1\columnwidth]{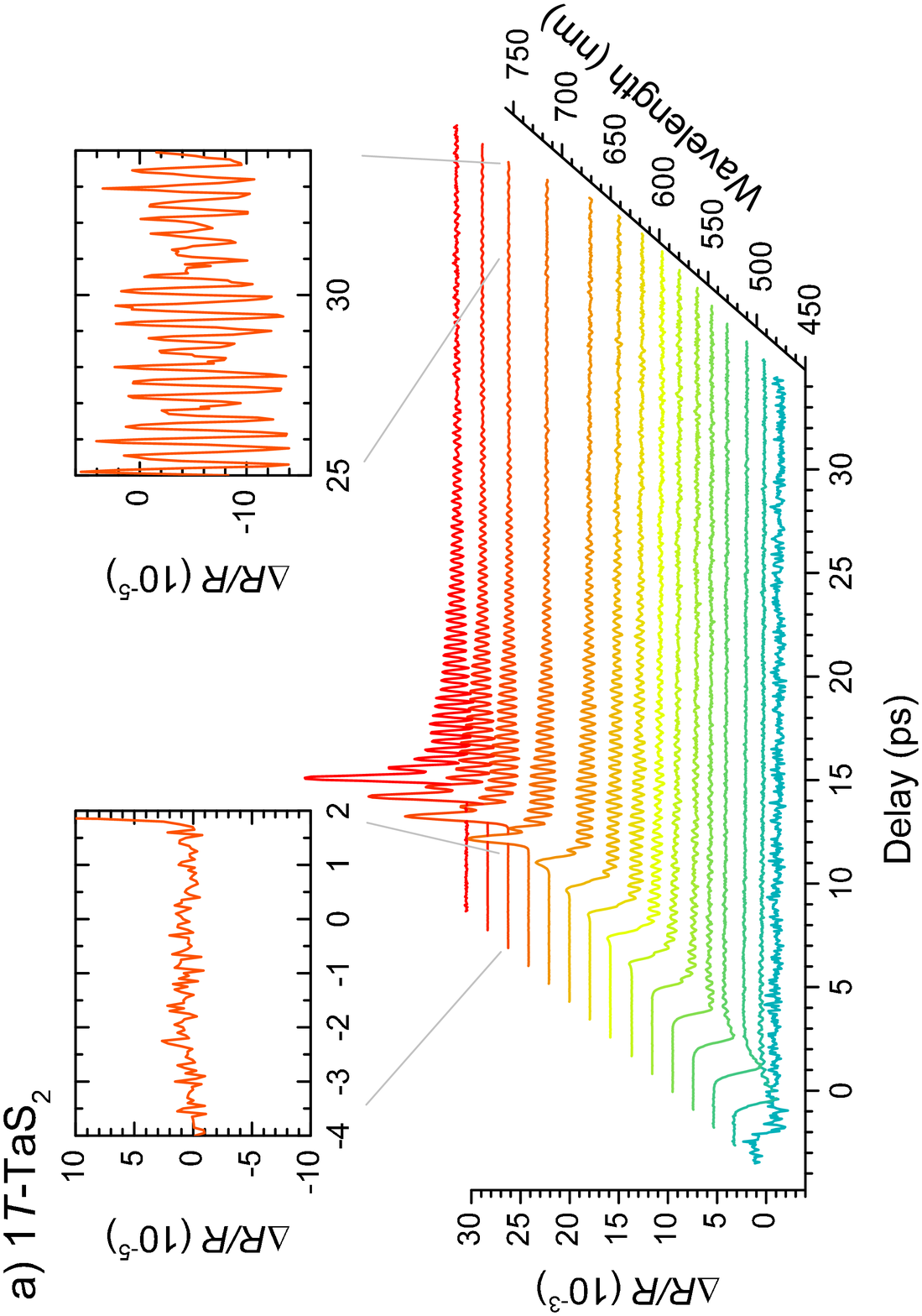}\medskip{}

\includegraphics[angle=-90,origin=lB,width=1\columnwidth]{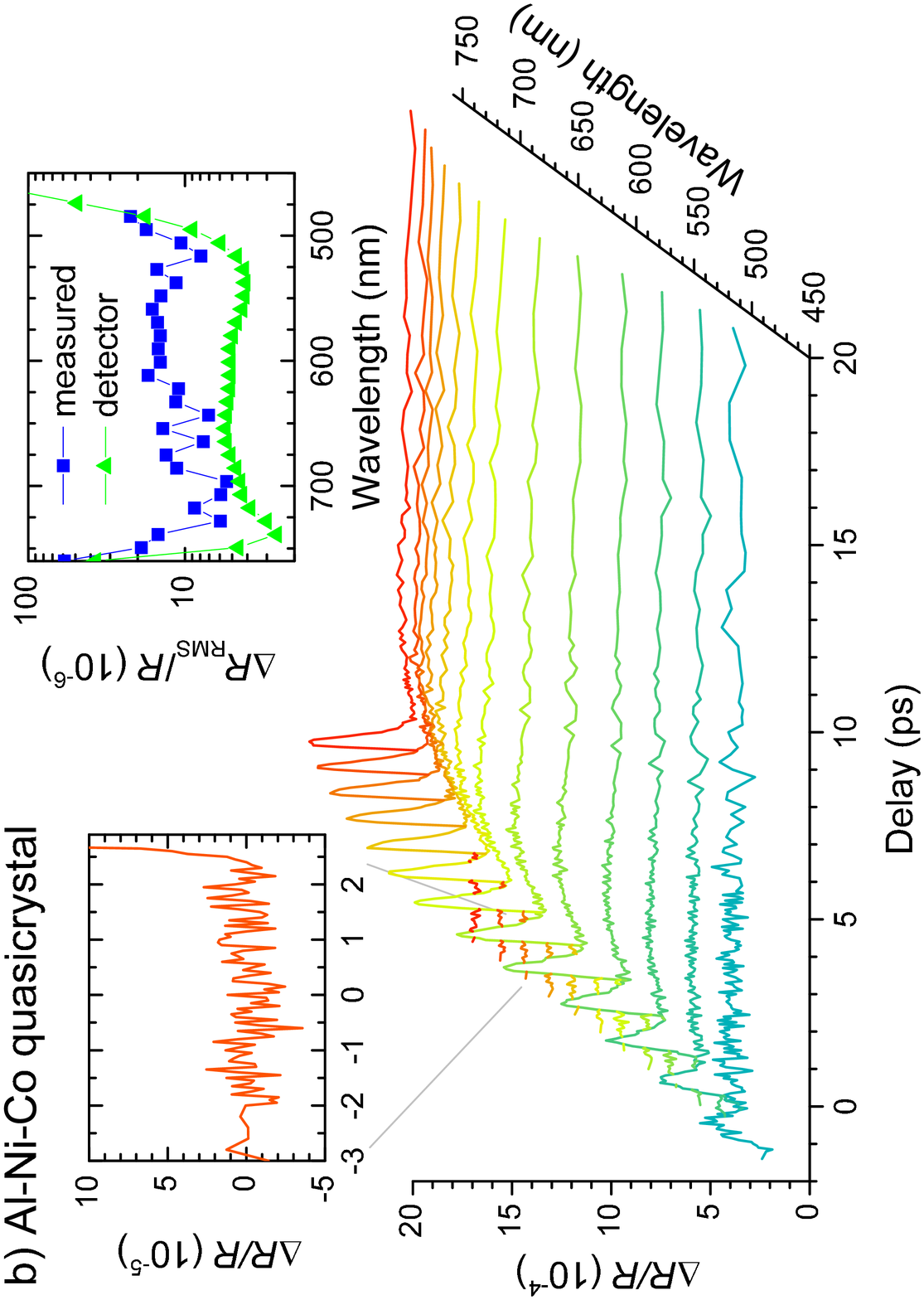}

\caption{a) The transient optical reflectivity of 1$T$-TaS$_{2}$ at 120 K.
The RMS noise level in the least noisy region is of order of 10$^{-5}$
as shown in the left panel so the weak coherent oscillations shown
in the right panel can be observed at long delays. b) The transient
optical reflectivity of an Al-Ni-Co quasicrystal\cite{MerteljOslak2009}
at room temperature. The noise level in the least noisy region is
shown in the left panel. The wavelength dependence of the RMS noise
level is shown in the right panel. The detector noise was estimated
from the unilluminated detector data. The chirp present in the supercontinuum
probe pulses can be clearly seen in both cases.}
\label{fig:example}
\end{figure}

The CW-illuminated-detector%
\footnote{A 20 W halogen lamp driven by a stabilized DC power supply was used
as a source.%
} noise at different modulation frequencies is shown in Fig. \ref{fig:noise}
(b). Due to the fixed number of readouts (2000) the total integration
time and consequently the total photelectron counts are inversely
proportional to the modulation frequency, $f$, resulting in a higher
signal noise floor at higher modulation frequencies. 

In Fig. \ref{fig:noise} (c) we show the illumination-intensity dependence
of the signal-noise floor at $f=3$ kHz. The measured signal-noise
floor is well described by the combination of the intensity-independent
dark-noise from Fig. \ref{fig:noise} (a) and the shot noise floor
shown in Fig. \ref{fig:noise} (c) with the blue dashed line. It can
be seen that the detector is capable of operation near the shot-noise
limit with the signal-noise floor of $\sim2\cdot10^{-6}$ at $\sim1$
s acquisition time.

In Fig. \ref{fig:example} we show two actual data examples measured
by the Si photodiode array, where the visible broadband supercontinuum
pulses were generated\cite{BradlerBaum2009} in a 2.5-mm thick sapphire
window from 800-nm \textbf{$\sim0.5$-}$\mu$J 50-fs pulses from a
high repetition rate optical regenerative amplifier. The IR part of
the supercontinuum ($\lambda>\sim750$ nm) was rejected by a combination
of Shott glass and interference filters. The reflected probe beam
was coupled to an in-house built 300-mm focal-length spectrometer
by means of a 0.6-mm core-diameter fiber. A 300 grooves/mm transmissive
optical grating was chosen to maximize throughput in the $800-400$
nm spectral region.

\begin{figure}
\bigskip{}
\includegraphics[width=0.6\columnwidth]{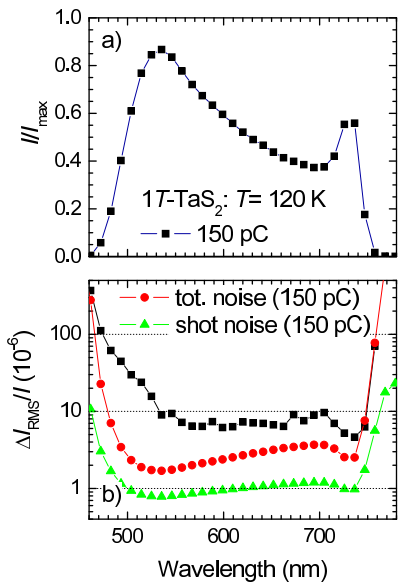}

\caption{(b) The RMS signal-noise floor for the case of coherent oscillations
measurement in $1T$-TaS$_{2}$. The full circles (red) and full triangles
(green) correspond to the estimated total detector noise and the shot
noise, respectively. The corresponding spectral dependence of the
photodiode current is shown in (a).}
\label{fig:noise-TaS2}

\end{figure}

The total integration time for each delay was $\sim$2 s resulting
in $\sim30$ min acquisition time for the case of the coherent oscillations\cite{DemsarForro2002}
in $1T$-TaS$_{2}$ {[}Fig. \ref{fig:example} (a){]} with $760$
delay points. Using 150 pC effective-well depth we obtained the RMS
noise floor of $<10^{-5}$ in the least noisy part of the spectrum
(see Fig. \ref{fig:noise-TaS2}). The most significant contribution
to the noise floor is clearly not due to the shot noise and the detector
noise but due to the supercontinuum source noise, which strongly increases
at shorter wavelengths.

A smaller effective-well depth of 50 pC was used for measurement in
an Al-Ni-Co quasi-crystal\cite{MerteljOslak2009} together with a
lower supercontinuum probe-pulse energy of $\sim170$ pJ per pulse,
resulting in $\sim2$ $\mu$J/cm$^{2}$ fluence with (100$\mu$m)$^{2}$
probe spot size. Due to the lower probe-pulse energy the noise floor
is somewhat larger ($\sim2\cdot10^{-5}$) as shown in Fig. \ref{fig:example}
(b).

\section{Conclusions}

A high signal-to-noise ($\sim$10$^{6}$) multichannel photodetector
for femtosecond transient optical spectroscopy in visible and NIR
was constructed from commercially available components. The 6-kHz
synchronous sampling rate enables lockin-like modulation detection
to suppress the optical source noise enabling $<10^{-5}$ transient
reflectivity/transmittance RMS signal-noise floor in a broad spectral
range as shown by two example measurements.
\begin{acknowledgments}
Work at Jozef Stefan Institute was supported by ARRS (Grant No. P1-0040). 
\end{acknowledgments}
\bibliographystyle{apsrev4-1}
\bibliography{biblio}
 
\end{document}